\documentclass[twocolumn,showpacs,preprintnumbers,amsmath,amssymb]{revtex4}
\usepackage{graphicx}
\usepackage{epsfig}
\usepackage{dcolumn}
\usepackage{bm}

\newcommand{\EE}{e^+e^-}
\newcommand{\MM}{\mu^+\mu^-}
\newcommand{\GG}{\gamma\gamma}

\newcommand{\ggee}{\gamma\gamma e^+e^-}
\newcommand{\gguu}{\gamma\gamma\mu^+\mu^-}
\newcommand{\ggll}{\gamma\gamma l^+l^-}

\newcommand{\jpsi}{J/\psi}
\newcommand{\ar}{\rightarrow}

\newcommand{\ptochic}{\psi(2S)\ar \gamma\chi_{c1,2}}
\newcommand{\ppjpsi}{\psi(2S)\ar\pi^0\pi^0 J/\psi}    
 
\newcommand{\bfg}{\begin{figure}}
\newcommand{\efg}{\end{figure}}
\newcommand{\bitm}{\begin{itemize}}
\newcommand{\eitm}{\end{itemize}}
\newcommand{\bnum}{\begin{enumerate}}
\newcommand{\enum}{\end{enumerate}}
\newcommand{\btbl}{\begin{table}}
\newcommand{\etbl}{\end{table}}
\newcommand{\btbu}{\begin{tabular}}
\newcommand{\etbu}{\end{tabular}}
\newcommand{\bcl}{\begin{center}}
\newcommand{\ecl}{\end{center}}

\newcommand{\beq}{\begin{equation}}
\newcommand{\eeq}{\end{equation}}
\newcommand{\beqr}{\begin{eqnarray}}
\newcommand{\eeqr}{\end{eqnarray}}
\newcommand{\J}{J/\psi}
\newcommand{\psip}{\psi(2S)}
\newcommand{\ppJ}{\pi^+\pi^-J/\psi}
\newcommand{\ra}{\rightarrow}
\newcommand{\Jpi}{\pi^0 J/\psi}
\newcommand{\Jeta}{\eta J/\psi}
\newcommand{\Jpipi}{\pi^0\pi^0 J/\psi}
\newcommand{\gxi}{\gamma\chi_{c1}}
\newcommand{\gx}{\gamma\chi_{c1,c2}}
\newcommand{\gxii}{\gamma\chi_{c2}}
\newcommand{\ggmm}{\gamma\gamma \mu^+\mu^-}

\newcommand{\MgJ}{M_{\gamma_h,J/\psi}}
\newcommand{\Mgg}{M_{\gamma\gamma}}
\begin{document}
\title{ \boldmath $\psi(2S)$ Decays into $\J$ plus Two Photons}
\author{J.~Z.~Bai$^1$,      Y.~Ban$^{10}$,      J.~G.~Bian$^1$,
X.~Cai$^{1}$,         J.~F.~Chang$^1$,       H.~F.~Chen$^{16}$,
H.~S.~Chen$^1$,       H.~X.~Chen$^{1}$,      J.~Chen$^{1}$,
J.~C.~Chen$^1$
% ...
\vspace{0.2cm}
\\(BES Collaboration)\\
\vspace{0.2cm}
$^1$ Institute of High Energy Physics, Beijing 100039, People's Republic of
     China\\
$^2$ China Center of Advanced Science and Technology, Beijing 100080,
     People's Republic of China\\
$^3$ Guangxi Normal University, Guilin 541004, People's Republic of China\\
$^4$ Henan Normal University, Xinxiang 453002, People's Republic of China\\
$^5$ Huazhong Normal University, Wuhan 430079, People's Republic of China\\
$^6$ Hunan University, Changsha 410082, People's Republic of China\\
 $^7$ Liaoning University, Shenyang 110036, People's Republic of China\\
$^{8}$ Nanjing Normal University, Nanjing 210097, People's Republic of
China\\
$^{9}$ Nankai University, Tianjin 300071, People's Republic of China\\
$^{10}$ Peking University, Beijing 100871, People's Republic of China\\
$^{11}$ Shandong University, Jinan 250100, People's Republic of China\\
$^{12}$ Shanghai Jiaotong University, Shanghai 200030,
        People's Republic of China\\
$^{13}$ Sichuan University, Chengdu 610064,
        People's Republic of China\\
$^{14}$ Tsinghua University, Beijing 100084,
        People's Republic of China\\
$^{15}$ University of Hawaii, Honolulu, Hawaii 96822\\
$^{16}$ University of Science and Technology of China, Hefei 230026,
        People's Republic of China\\
$^{17}$ Wuhan University, Wuhan 430072, People's Republic of China\\
$^{18}$ Zhejiang University, Hangzhou 310028, People's Republic of China\\
\vspace{0.4cm}
$^{\ast}$ Visiting professor to University of Michigan, Ann Arbor, MI
48109 USA}
%\vspace{0.4cm}
 
%$^{\ast}$ 
%Visiting professor to University of Michigan, Ann Arbor, MI 48109 USA}
%\end{center}

\begin{abstract}
Using $\gamma \gamma J/\psi, J/\psi \ra e^+ e^-$ and $\mu^+ \mu^-$
events from a sample of $14.0\times 10^6$ $\psip$ decays collected
with the BESII detector, the branching fractions for $\psip\ra
\pi^0\J$, $\eta\J$, and
$\psi(2S)\ar\gamma\chi_{c1},\gamma\chi_{c2}\ar\gamma\gamma\jpsi$ are
measured to be $B(\psip\ra \pi^0\J) = (1.43\pm0.14\pm0.13)\times
10^{-3}$, $B(\psip\ra \eta\J) = (2.98\pm0.09\pm0.23)\%$,
$B(\psi(2S)\ar\gamma\chi_{c1}\ar\gamma\gamma\jpsi) = (2.81\pm0.05\pm
0.23)\%$, and $B(\psi(2S)\ar\gamma\chi_{c2}\ar\gamma\gamma\jpsi) =
(1.62\pm0.04\pm 0.12)\%$.
 
\vspace{1pc}
\end{abstract}
\maketitle

\normalsize

\section{Introduction}

Experimental data for the processes $\psip\ra\pi^0\J$, $\eta\J$, and
$\gamma\chi_{c1,2}$ are scarce and were mainly taken in the 1970s and
80s \cite{MK1,CNTR1,CNTR2,DASP,MK2,CB}.  The branching fractions from
different experiments do not agree well, and the $\Jpi$ channel is
measured with low precision.  In particular, improved branching
fractions for $\psip\ra\gamma\chi_{cJ}$ are very important for the
measurement of $\chi_{cJ}$ decay branching fractions using $\psip$
data.  All these appeal for high statistics measurements of these
channels.

In this paper, we report on the analysis of $\psip\ra\pi^0\J$,
$\eta\J$ and $\gamma\chi_{c1,2}$ decays based on a sample of
$14.0\times 10^6$ $\psip$ events collected with the BESII detector.
% at a center-of-mass energy corresponding to $M_{\psip}$. 
The first two decays are important to test various theoretical
predictions for the ratios
$R=\frac{\Gamma(\psip\ra\Jpi)}{\Gamma(\psip\ra\Jeta)}$, $R^{\prime}=\frac{\Gamma(\Upsilon^\prime\ra\eta\Upsilon)}
{\Gamma(\psip\ra\Jeta)}$, and $R^{\prime\prime}=\frac{\Gamma(\Upsilon^{\prime\prime}\ra\eta\Upsilon)}
{\Gamma(\psip\ra\Jeta)}$.
The ratio
$R$ has been
calculated by different theoretical
approaches~\cite{Ioffe},~\cite{Miller} ~\cite{aldo}, and the ratios
$R^{\prime}$ and  $R^{\prime\prime}$
have been predicted in the framework of the QCD
multipole expansion mechanism ~\cite{Yan}~\cite{kuang}.
% These
%predictions could be tested by our measurements.
 
\section{The BES detector}
The Beijing Spectrometer (BES) detector is a conventional solenoidal
magnet detector that is described in detail in Ref.~\cite{bes}; BESII
is the upgraded version of the BES detector~\cite{bes2}. A 12-layer
vertex chamber (VC) surrounding the beam pipe provides trigger
information. A forty-layer main drift chamber (MDC), located radially
outside the VC, provides trajectory and energy loss ($dE/dx$)
information for charged tracks over $85\%$ of the total solid angle.
The momentum resolution is $\sigma _p/p = 0.017 \sqrt{1+p^2}$ ($p$ in
$\hbox{\rm GeV}/c$), and the $dE/dx$ resolution for hadron tracks is
$\sim 8\%$.  An array of 48 scintillation counters surrounding the MDC
measures the time-of-flight (TOF) of charged tracks with a resolution
of $\sim 200$ ps for hadrons.  Radially outside the TOF system is a 12
radiation length, lead-gas barrel shower counter (BSC).  This measures
the energies of electrons and photons over $\sim 80\%$ of the total
solid angle with an energy resolution of $\sigma_E/E=22\%/\sqrt{E}$
($E$ in GeV).  Outside of the solenoidal coil, which provides a
0.4~Tesla magnetic field over the tracking volume, is an iron flux
return that is instrumented with three double layers of counters that
identify muons of momentum greater than 0.5~GeV/$c$.

A GEANT3 based Monte Carlo program with
detailed consideration of the detector performance (such as dead
electronic channels) is used to simulate the BESII detector.  The
consistency between data and Monte Carlo has been carefully checked in
many high purity physics channels, and the agreement is quite
reasonable.

%%\section{Monte Carlo}

\section{Event selection}
 The data sample used for this analysis consists of $(14.00\pm
 0.56)\times10^6$ $\psip$ events ~\cite{Npsip} collected with the
 BESII detector at the center-of-mass energy $\sqrt s=M_{\psip}$.  The
 channels investigated are $\psip$ decaying into $(\pi^0,\eta)\J$ and
 $\gamma\chi_{c1,2}$, with $\chi_{c1,2}$ decaying to $\gamma\J$,
 $\pi^0$ and $\eta$ to two photons, and $\J$ to lepton pairs. They all have
 a final $\gamma\gamma l^+l^-$ ($l=e,\mu$) event topology.

\subsection{\boldmath General selection for $\gamma\gamma l^+l^-$ events} 

A neutral cluster is considered to be a photon candidate if it is
located within the BSC fiducial region ($|\cos\theta|<0.75$), the
energy deposited in the BSC is greater than 50 MeV, the first hit
appears in the first 6 radiation lengths, the angle between the
cluster and the nearest charged track is greater than $14^\circ$.
Each charged track is required to be well fit by a three-dimensional
helix and to have a polar angle, $\theta$, within the fiducial region
$|\cos\theta|<0.8$. To ensure tracks originate from the interaction
region, we require $V_{xy}=\sqrt{V_x^2+V_y^2}<2$ cm and $|V_z|<20$ cm,
where $V_x$, $V_y$, and $V_z$ are the x, y, and z coordinates of the
point of closest approach of each track to the beam axis.

Events with two charged tracks and two or three photon
candidates are subject to further selection criteria. The two charged
tracks are identified as an electron pair or muon pair
if $0<S<0.6$ or $S>0.9$, respectively, where
$$S=\sqrt{(\frac{E_{sc1}}{p_1}-1)^2+(\frac{E_{sc2}}{p_2}-1)^2}$$
and $p$ and $E_{sc}$ are the momentum and the deposited energy in the 
BSC 
of a charged track.

A five constraint (5C) kinematic fit to the hypothesis $\psip\ra\ggll$
with the invariant mass of the lepton pair constrained to the $\J$ mass is
performed, and the fit probability is required to be greater than
0.01.  For events with three photon candidates, the combination of two
photons having the smaller $\chi^2$ is chosen.  Fig. 1 shows a scatter
plot of the invariant mass of the reconstructed $\J$ and the photon
with higher energy ($\MgJ$) versus the invariant mass of two photons
($\Mgg$) for the $\ggmm$ final state.  The corresponding plot for the
$\ggee$ state is very similar.  The $\eta$, $\chi_{c1}$, and
$\chi_{c2}$ signals are quite prominent, while the $\pi^0$ signal is
much less so.  The corresponding plot for 200,000 Monte Carlo
simulated $\psip\ra\Jpipi$ events, which is the main background for
the studied channels, is shown in Fig. 2.
\begin{figure}[htbp]
%\centerline{\includegraphics[height=5.5cm,width=7cm]{/data2/wangzy/eps/scatter.eps}}
\centerline{\includegraphics[height=5.5cm,width=7cm]{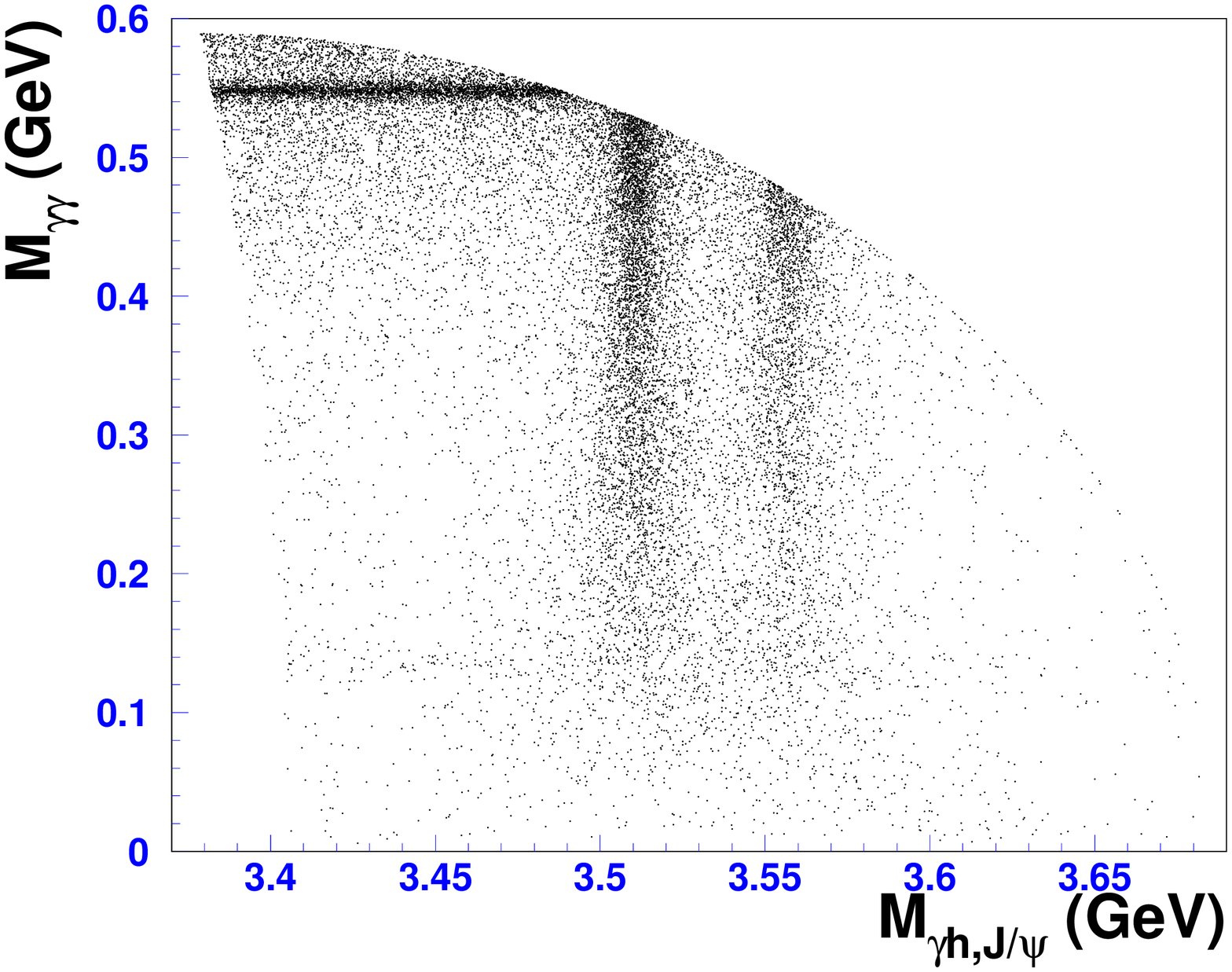}}
\caption{\label{fig:MggDT}
   $\MgJ$ versus $\Mgg$ after general selection of $\gguu$ events.}
\end{figure}
\begin{figure}[htbp]
%\centerline{\includegraphics[height=5.5cm,width=7cm]{/data2/wangzy/eps/ppbg.eps}}
\centerline{\includegraphics[height=5.5cm,width=7cm]{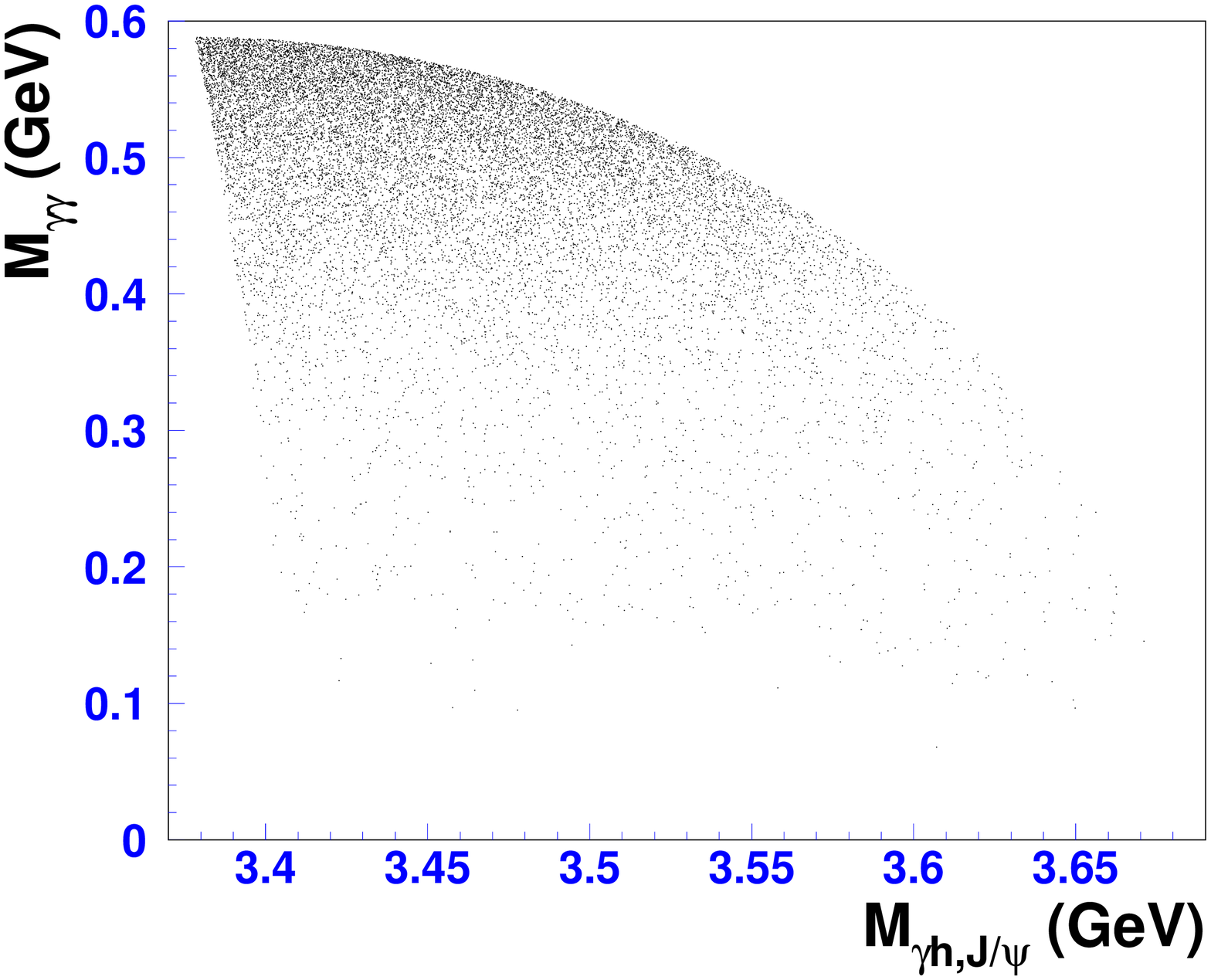}}
\caption{\label{fig:MggMC} 
   $\MgJ$ versus $\Mgg$ after general selection for 200,000 $\psip\ra\Jpipi$ 
   Monte Carlo events ($\gguu$ final state).} 
\end{figure}

 \subsection{\boldmath Selection of $\psip\ra\Jpi$}  
 
 To remove the huge background from $\psip\ra\gx$ under the
 $\psip\ra\Jpi$ signal, we require $\MgJ$ to be less than 3.49 or greater
 than 3.58 GeV/c$^2$.  Fig. 3 shows, after this requirement, the
 distribution of invariant mass, $\Mgg$, where the smooth background
 is contributed by $\ptochic$ and $\ppjpsi$. A Breit Wigner with a
 double Gaussian mass resolution function to describe the $\pi^0$
 resonance plus a third-order background polynomial is fitted to the
 data. The fit gives
  $N_{\EE}^{\pi^0}=123\pm 18$ for the $\ggee$ state and
   $N_{\MM}^{\pi^0}=155\pm 20$ for the $\ggmm$ state.   
   In the fit, the mass resolution and the area ratio of the two
 Gaussians are fixed to the values determined by the Monte Carlo.  The
 fit is also performed with a background function determined by Monte
 Carlo simulated $\ppjpsi$ and $\ptochic$ events that satisfy the
 selection criteria, where the two processes are weighted according to
 their branching fractions.  The differences in the number of events
 obtained with the two backgrounds (5.1\% for $\ggee$ and 4.3\% for
 $\gguu$ ) are included in the systematic errors. 

\begin{figure}[htbp]
\hspace{0.2cm}
%\centerline{\includegraphics[height=4.5cm,width=8cm]{/data2/wangzy/eps/pi0fit.eps}}
\centerline{\includegraphics[height=4.5cm,width=8cm]{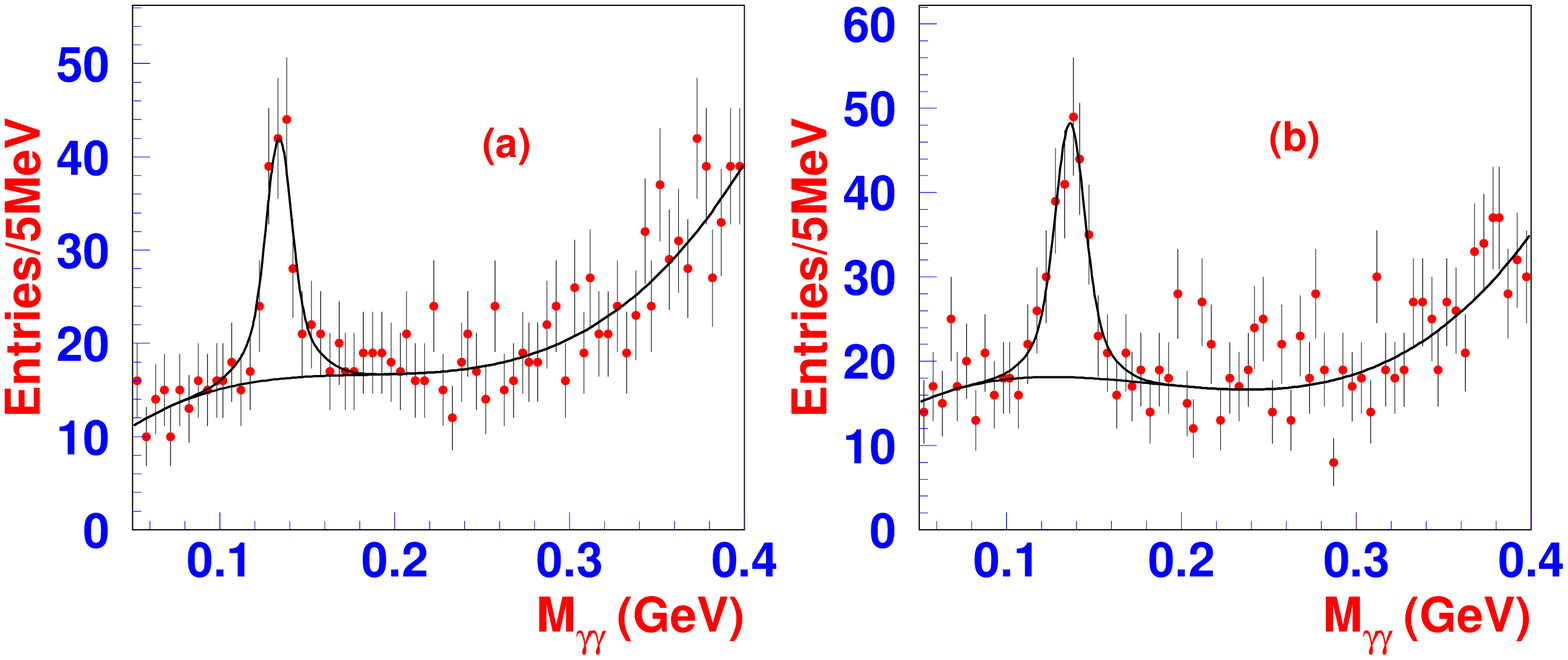}}
\caption{\label{fig:MJpiee}
  Two photon invariant mass distribution for candidate $\psip\ra\Jpi$ 
events for (a) $\ggee$ and (b) $\gguu$.}
\end{figure}

\subsection{\boldmath Selection of $\psip\ra\Jeta$}

In this channel the main backgrounds are from $\psip\ra\Jpipi$ and
$\gx$. By requiring $\MgJ<3.49$ GeV/c$^2$, most background from
$\psip\ra \gx$ is removed.  The resultant plot, shown in Fig. 4, shows a
clear $\eta$ signal superimposed on background, mainly from
$\psip\ra\Jpipi$.  A fit is made using a Breit-Wigner resonance
convoluted with a mass resolution function for the $\eta$ signal plus
a polynomial background, where the width of the $\eta$ is fixed to its
Particle Data Group (PDG)
value ~\cite{PDG} and the background function is determined from
$\psip\ra\Jpipi$ Monte Carlo simulated events that satisfy the same criteria as
the data. The fit yields
$N_{\EE}^{\eta}=2465\pm 101$ for the $\ggee$ state and
$N_{\MM}^{\eta}=3290\pm 148$ for the $\ggmm$ state.
The fitted values of the $\eta$ mass are $547.6\pm 0.1$ MeV/c$^2$ for
the $\ggee$ channel and $547.7\pm 0.1$ MeV/c$^2$ for the $\gguu$ channel,
consistent with the PDG value within $2\sigma$.

A fit using a fourth-order background polynomial with parameters free is
also performed to estimate the systematic error due to the
background shape.  This error is negligibly small.  
\begin{figure}[htbp]
\hspace{0.2cm}
%\centerline{\includegraphics[height=4.cm,width=7.5cm]{/data2/wangzy/eps/etafit.eps}}
\centerline{\includegraphics[height=4.cm,width=7.5cm]{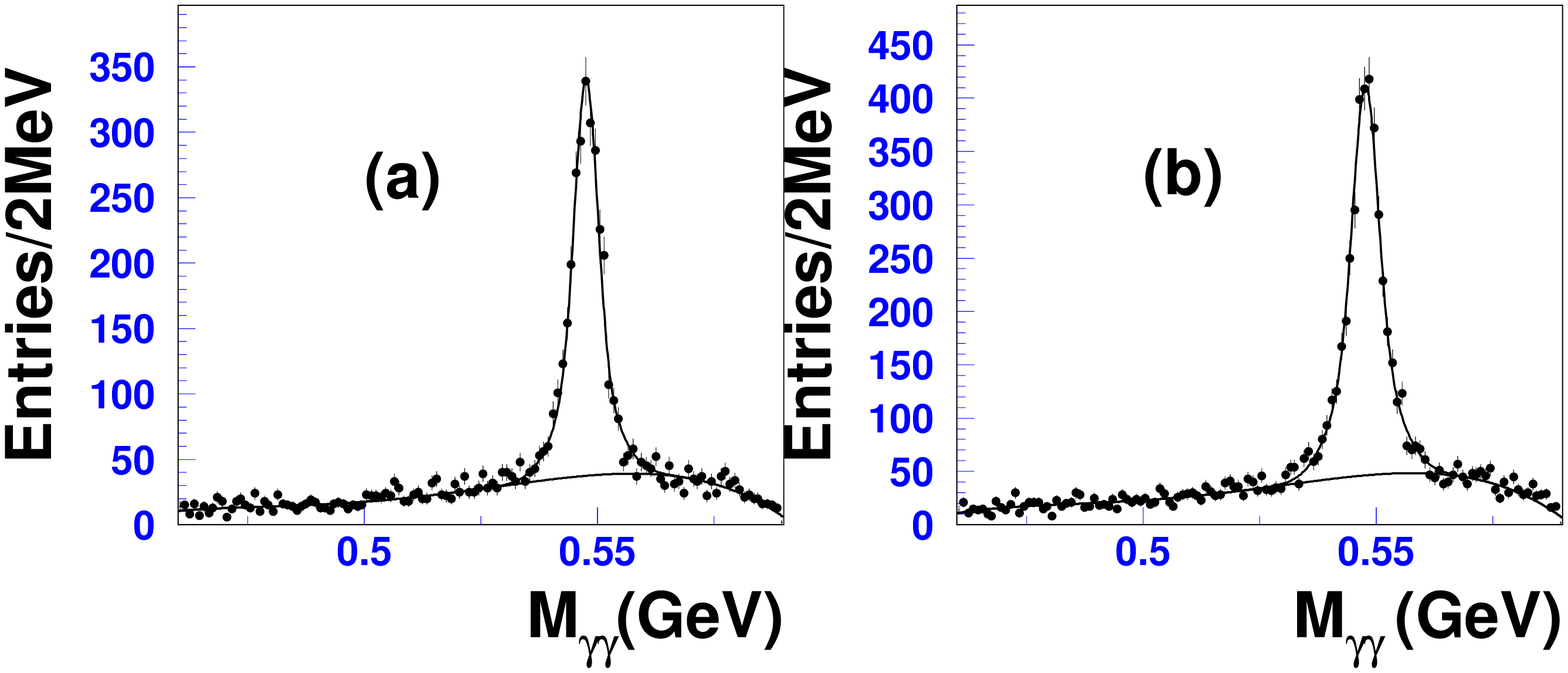}}
\caption{\label{fig:MJetaee}
  Two photon invariant mass distribution for candidate $\psip\ra\Jeta$
events for (a) $\ggee$ and (b) $\gguu$.}
\end{figure}

\subsection{\boldmath Selection of $\psip\ra\gx$}

   The processes $\psip\ra\Jpi$, $\Jeta$, and $\Jpipi$ contribute to
the background for this channel. By requiring $\Mgg<0.53$ GeV/c$^2$, most of
the background from $\psip\ra\Jeta$ and a significant portion from
$\psip\ra\Jpipi$ are rejected.  Fig. 5 shows the
$\MgJ$ distribution for candidate $\psip\ra\gx$ events.  The
remaining background is mainly due to $\psip\ra\Jpipi$.  The
contribution from $\psip\ra\Jpi$ is negligible due to its tiny
branching fraction.  Fig. 6 shows the $\MgJ$ distribution for
$\psip\ra\Jpipi$ Monte Carlo simulated events before and after the
$\Mgg<0.53$ GeV/c$^2$ requirement, the latter one is taken as the background shape
in the fit.  Two Breit-Wigner resonances convoluted with mass
resolution functions plus a background function are fitted to the data.
The widths for the $\chi_{c1,2}$ are fixed to the PDG values, and the mass
resolution functions are determined by Monte Carlo simulation.  The
fit yields:
$$N_{\EE}^{\chi_{c1}}=5263\pm124, \:  N_{\EE}^{\chi_{c2}}=2512\pm82,$$
$$N_{\MM}^{\chi_{c1}}=6752\pm178, \:  N_{\MM}^{\chi_{c2}}=3358\pm96,$$
with the fitted masses of $\chi_{c1}$ and $\chi_{c2}$ equal to
$3510.9\pm 1.0$ MeV/c$^2$ and $3555.9\pm 1.0$ MeV/c$^2$, respectively, consistent
with the PDG values. 

\begin{figure}[ht]
\hspace{0.3cm}
%\centerline{\includegraphics[height=4cm,width=8.5cm]{/data2/wangzy/eps/chicfit.eps}}
\centerline{\includegraphics[height=4cm,width=8.5cm]{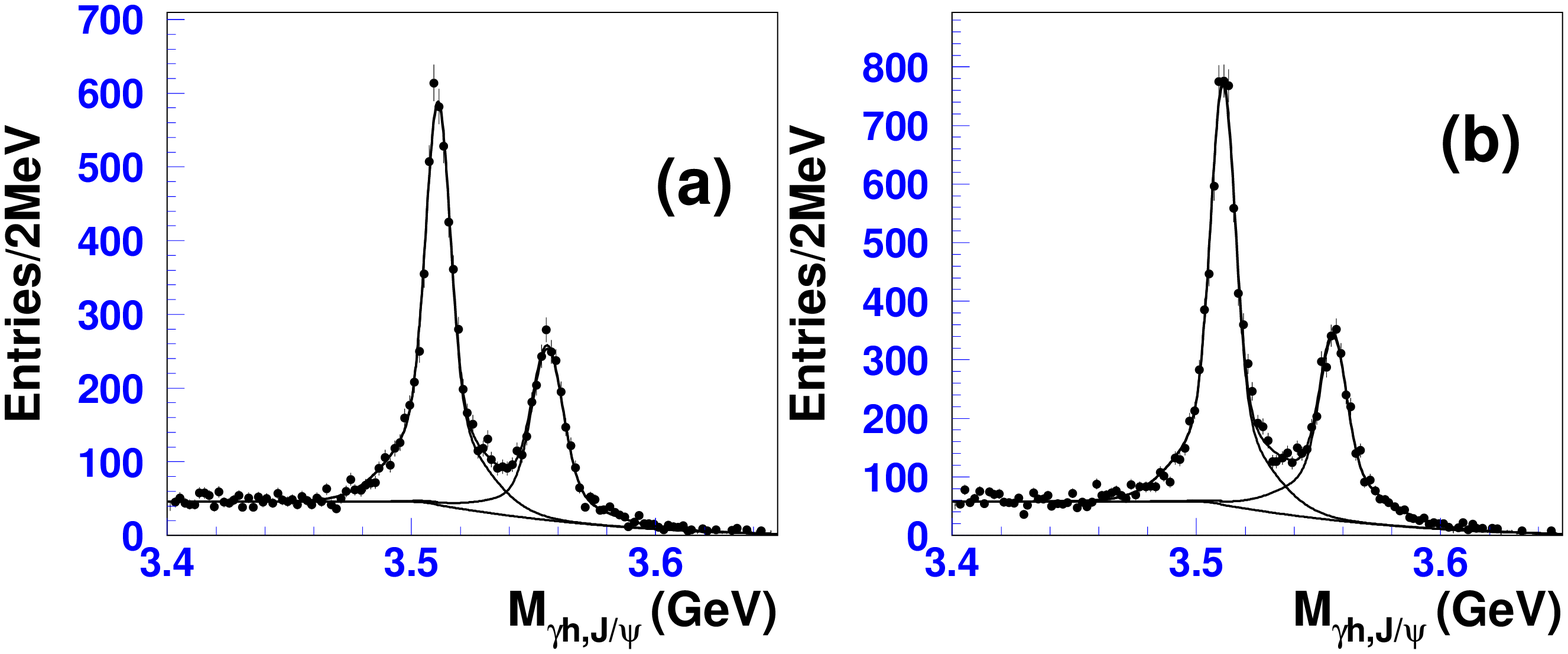}}
\caption{\label{fig:MgJee}
  Invariant mass $\MgJ$ distribution for candidate $\psip\ra\gx$ 
events for (a) $\ggee$ and (b) $\gguu$.}
\end{figure}

\begin{figure}[hb]
\hspace{0.2cm}
%\centerline{\includegraphics[height=4cm,width=8.5cm]{/data2/wangzy/eps/p0p0bg.eps}}
\centerline{\includegraphics[height=4cm,width=8.5cm]{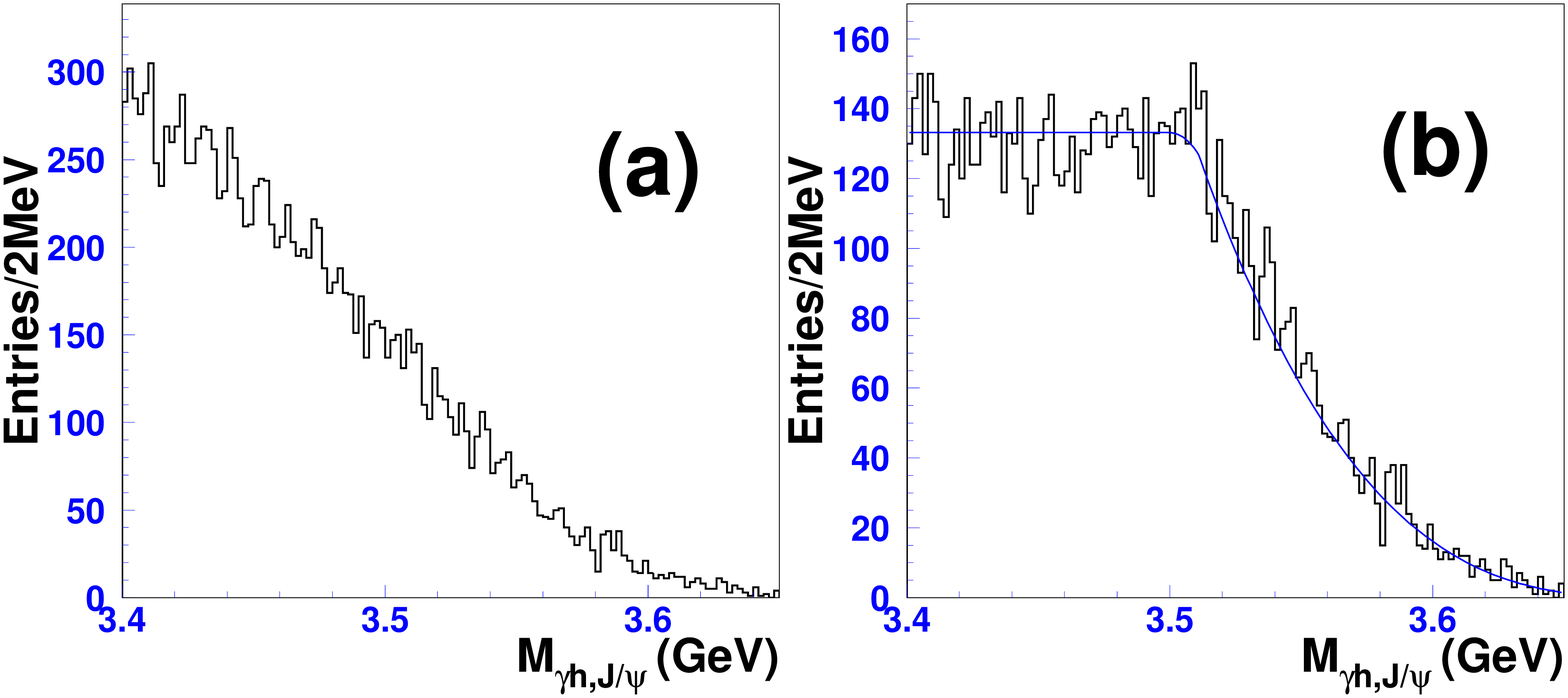}}
\caption{\label{fig:MgJMCb} 
  Invariant mass $\MgJ$ distribution for Monte Carlo simulated  $\psip\ra\Jpipi$ 
events ($\gguu$ final state). (a) Before the $\Mgg<0.53$ GeV/c$^2$ requirement. 
(b) After the $\Mgg<0.53$ GeV/c$^2$ requirement.}
\end{figure}

\section{Branching fraction determination}

  For $\psip\ra X$, the  branching fraction is determined from 
$$B(\psip\ra X)=$$
$$\frac{n^{obs}(\psip\ra X\ra Y)}
  {N_{\psip}\cdot B(X\ra Y)\cdot\epsilon(\psip\ra X\ra Y)},$$
where Y stands for the final state, X the intermediate state,
and $\epsilon$ the detection efficiency. The branching fraction of   
$X\ra Y$ is taken from the PDG. 

 \subsection{Detection efficiency}
 
    The detection efficiency is the product of the trigger efficiency
$\epsilon_{trg}$ and the reconstruction-selection efficiency
$\epsilon_{rs}$.  For the BES detector, the trigger efficiency for
hadronic events is $1.000\pm 0.005$~\cite{R}. The
reconstruction-selection efficiency is determined by Monte Carlo
simulation.
For the signal channels studied, generators
taking into account phase space, angular distributions, and final
state radiation are used for the event simulations.  For the channel
$\psip\ra\Jpipi$, the common background for all signals, we use a
generator, which gives the correct dipion mass and angular
distributions ~\cite{gener}. 

For each of the channels analyzed, 50,000 Monte Carlo events are
subjected to the same reconstruction and event selection as used for the
data to determine the detection efficiencies, which are listed in
Table~\ref{results}.

\subsection{Efficiency corrections and systematic errors} 

Because the Monte Carlo is imperfect, it is
necessary to correct the detection efficiencies obtained from
simulations for the differences between the Monte Carlo simulation and
the data.
Differences come from the efficiencies of MDC tracking, particle
identification, photon identification, and kinematic fitting.  In
addition, the uncertainties of the background shapes (estimated in
Section 3), the number of $\psip$ events, and the branching fractions of
the intermediate states also contribute to the final systematic error.

To investigate the difference in the lepton track efficiencies of the
Monte Carlo simulation and the data, the lepton pair sample from the
decay $\psip\ra\ppJ, \J\ra l^+l^-$, which closely simulates the
behavior of the lepton pair in the channels under study, is used. This
study finds the tracking efficiency correction factor is
$1.012\pm0.009$ for $\EE$ pairs and $1.002\pm0.008$ for $\MM$ pairs.
For charged particle identification, $S$ is used to separate $\EE$ and
$\MM$ pairs. The same lepton pair sample is used to determine the
particle identification efficiency difference between Monte Carlo
simulation and data by determining efficiences for each with and
without this particle identification requirement.  The correction
factor is found to be $0.951\pm0.008$ for $\EE$ pairs and
$0.972\pm0.006$ for $\MM$ pairs.

For the photon selection used, studies show that the efficiency
difference between data and Monte Carlo is 2\% for each photon
\cite{photons}. We take this difference as the systematic error in
photon selection, and no correction to the efficiency is made. In
addition, the rib in the BSC causes an inefficiency in photon
detection. The systematic error due to the rib efficiency, listed in
table~\ref{factor}, is obtained by comparing results with photons in
the rib region removed with those when they are not removed.

The systematic error due to kinematic fitting comes from the
differences between data and Monte Carlo simulation in the
measurements of track momentum, the track fitting error matrix, and
the photon energy and direction.  For the charged track part, the
difference is estimated using the $\psip\ra\ppJ,\J\ra l^+l^-$
sample. For the photon part, a careful calibration of the neutral
cluster information in the BSC is performed and the difference with
and without the calibration applied to both the
data and Monte Carlo is used to determine the systematic error in this
part~\cite{lambda}. 

  Table~\ref{factor} summarizes the efficiency correction factors and uncertainties from all sources, while
Table~\ref{totsys} lists the systematic errors for the channels under
study. The branching fractions and corresponding
errors for all intermediate state decays are taken from the
PDG~\cite{2003pdg}.
%%%%%%%%%%%%%%%%%  table of correction factor  %%%%%%%%%%%%%%%%%
\begin{table*}[htpb]
\doublerulesep 0.5pt
\begin{center}
\caption{\label{factor} Efficiency correction factors.}
\vskip 0.2 cm
%\begin{center}
\begin{tabular}{c|cc|cc}              \hline        \hline
Channel&\multicolumn{2}{|c|}{$\pi^0\jpsi$}&\multicolumn{2}{|c}{$\eta\jpsi$}\\\hline
Final state&$\ggee$&$\gguu$&$\ggee$&$\gguu$\\\hline
Track selection&$1.012\pm 0.009$&$1.002\pm 0.008$&$1.012\pm
0.009$&$1.002\pm 0.008$\\
Particle ID&$0.951\pm 0.008$&$0.972\pm 0.006$&$0.951\pm
0.008$&$0.972\pm 0.006$\\
5-C fit&$1.000\pm 0.014$&$1.00\pm 0.02$&$1.000\pm 0.016$&$1.000\pm
0.038$\\
$\gamma$ efficiency&$1.00\pm 0.04$&$1.00\pm 0.04$&$1.00\pm
0.04$&$1.00\pm 0.04$\\
BSC rib&$1.000\pm 0.023$&$1.000\pm 0.034$&$1.000\pm 0.031$&$1.000\pm
0.036$\\\hline
Total correction&$0.962\pm0.050$&$0.974\pm0.057$&$0.962\pm0.055$
&$0.974\pm0.067$\\\hline\hline\hline
Channel&\multicolumn{2}{|c|}{$\gamma\chi_{c1}$}&\multicolumn{2}{|c}
{$\gamma\chi_{c2}$}\\\hline
Final state&$\ggee$&$\gguu$&$\ggee$&$\gguu$\\\hline
Track selection&$1.012\pm 0.009$&$1.002\pm 0.008$&$1.012\pm
0.009$&$1.002\pm 0.008$\\
Particle ID&$0.951\pm 0.008$&$0.972\pm 0.006$&$0.951\pm
0.008$&$0.972\pm 0.006$\\
5-C fit&$1.000\pm 0.015$&$1.000\pm 0.049$&$1.000\pm 0.018$&$1.00\pm
0.052$\\
$\gamma$ efficiency&$1.00\pm 0.04$&$1.00\pm 0.04$&$1.00\pm
0.04$&$1.00\pm 0.04$\\
BSC rib&$1.000\pm 0.043$&$1.000\pm 0.040$&$1.000\pm 0.019$&$1.000\pm
0.024$\\\hline
Total correction&$0.962\pm0.061$&$0.974\pm0.075$&$0.962\pm0.049$
&$0.974\pm0.070$\\\hline
\end{tabular}
\end{center}
\end{table*}
%%%%%%%%%%%%%%%%%5
\begin{table*}[htpb]
\doublerulesep 0.5pt
\begin{center}
\caption{\label{totsys} Systematic errors (\%)}
\vskip 0.2 cm
%\begin{center}
\begin{tabular}{c|cc|cc}              \hline        \hline
Channel&\multicolumn{2}{|c|}{$\pi^0\jpsi$}&\multicolumn{2}{|c}{$\eta\jpsi$}\\\hline
Final state&$\ggee$&$\gguu$&$\ggee$&$\gguu$\\\hline
efficiency correction&6.3&5.9&5.7&6.9\\
Number of $\psi(2S)$ events&4&4&4&4\\
${\cal B}(\pi^0,\eta\ar\GG)$&negligible&negligible&0.65&0.65\\
${\cal B}(\jpsi\ar\EE,\MM)$&1.7&1.7&1.7&1.7\\
background shape&5.1&4.3&negligible&negligible\\\hline\hline
Total systematic error (\%)&9.20&8.50&7.20&8.18\\\hline\hline
Channel&\multicolumn{2}{|c|}{$\gamma\chi_{c1}$}&\multicolumn{2}{|c}
{$\gamma\chi_{c2}$}\\\hline
Final state&$\ggee$&$\gguu$&$\ggee$&$\gguu$\\\hline
efficiency correction&6.3&7.7&5.1&7.2\\
Number of $\psi(2S)$ events&4&4&4&4\\
${\cal B}(\chi_{cJ}\ar\gamma\jpsi)$&8.5&8.5&8.9&8.9\\
${\cal B}(\jpsi\ar\EE,\MM)$&1.7&1.7&1.7&1.7\\\hline\hline
Total systematic error (\%)&11.44&12.26&11.14&12.25\\\hline
\end{tabular}
\end{center}
\end{table*}

\subsection{Results and discussion}

%%%%%%%%%%%%%%%%%%%%%  table of Branching ratio  %%%%%%%%%%%%
%\begin{sidewaystable*}[htb]
\begin{table*}[htbp]
\doublerulesep 0.5pt
\caption{\label{results} Results.  Note that much of the systematic
  error on $B(\psi(2S) \ra \gamma \chi_{c1,c2})$ is due to the
  uncertainty on $B(\chi_{c1,c2} \ra \gamma J/\psi)$.  }
\vskip 0.2 cm
\begin{center}
{\footnotesize{
\begin{tabular}{c|cc|cc}              \hline        \hline
Channel&\multicolumn{2}{c}{$\pi^0\jpsi$}&\multicolumn{2}{c}{$\eta\jpsi$}\\\hline
Final state&$\ggee$&$\gguu$&$\ggee$&$\gguu$\\\hline
Number of events&$123\pm 18$&$155\pm 20$&$2465\pm 101$&$3290\pm 148$\\
Efficiency($\%$)&11.21&13.34&26.94&34.07\\
Sys. error (\%)&9.68&8.77&8.54&8.40\\
Correction factor&0.962&0.974&0.962&0.974\\\hline
BR (\%)&$0.139\pm 0.020\pm 0.013$&$0.147\pm 0.019\pm
0.013$&$ 2.91\pm 0.12\pm 0.21$&$3.06\pm 0.14\pm 0.25 $\\\hline
Combine BR  (\%)&\multicolumn{2}{|c|}{$0.143\pm 0.014\pm 0.013$}
&\multicolumn{2}{|c}{$2.98\pm 0.09\pm 0.23$}\\
PDG (\%)&\multicolumn{2}{|c|}{$0.096\pm 0.021 $}&\multicolumn{2}{|c}
{$3.13\pm 0.21$}\\\hline\hline\hline
Channel&\multicolumn{2}{c}{$\gamma\chi_{c1}$}&\multicolumn{2}{c}
{$\gamma\chi_{c2}$}\\\hline
Final state&$\ggee$&$\gguu$&$\ggee$&$\gguu$\\\hline
Number of events&$5263\pm 124$&$6752\pm 178$&$2512\pm 82$&$3358\pm 96$\\
Efficiency($\%$)&23.88&29.24&19.70&25.54\\
Sys. error (\%)&12.23&12.45&12.10&12.44\\
Correction factor&0.962&0.974&0.962&0.974\\\hline
BR (\%)&$8.73\pm 0.21\pm 1.00$&$9.11\pm 0.24\pm 1.12$&
$7.90\pm 0.26\pm 0.88$&$8.12\pm 0.23\pm 0.99$\\
Combine BR (\%)&\multicolumn{2}{|c|}{$8.90\pm 0.16\pm
1.05$}&\multicolumn{2}{|c}{$8.02\pm 0.17\pm 0.94$}\\
PDG (\%)&\multicolumn{2}{|c|}{$8.4\pm 0.6$}&\multicolumn{2}{|c}
{$6.4\pm 0.6$}\\\hline
\end{tabular}
}}
\end{center}
%\end{sidewaystable*}
\end{table*}

Using the fitting results and the efficiencies and 
correction factors for each channel, we determine the branching fractions 
listed in Table~\ref{results}. We also obtain the product branching fractions
\beqr
B(\psip\ra\gxi)\cdot B(\chi_{c1}\ra\gamma\J)=\nonumber\\
(2.81\pm 0.05\pm 0.23)\%\nonumber,
\eeqr
\beqr
B(\psip\ra\gxii)\cdot B(\chi_{c2}\ra\gamma\J)=\nonumber\\
(1.62\pm 0.04\pm 0.12)\%~\nonumber.
\eeqr

Our $B(\psip\ra\Jpi)$ measurement has improved precision
by more than a factor of two compared with other 
experiments, and the BES $\psip\ra\Jeta$ branching fraction 
is the most accurate single measurement.  Our
$B(\psip\ra\Jpi)$ agrees better with the Mark-II result \cite{MK2} than with the 
Crystal Ball result \cite{CB}, while $B(\psip\ra\gx)$ agrees well with the Crystal 
Ball results \cite{CB}.  Much of the systematic error on
$B(\psip\ra\gx)$ comes from the uncertainties on $B(\chi_{c1}\ra\gamma\J)$
and  $B(\chi_{c2}\ra\gamma\J)$.

Using Partially Conserved Axial-vector Currents (PCAC), Miller
$et~al$~\cite{Miller} predicts:
\beq\label{eq1}
R=\frac{\Gamma(\psip\ra\Jpi)}{\Gamma(\psip\ra\Jeta)} = \frac{27}{16}(\frac{p_\pi}{p_\eta})^3 r^2,
\eeq
where $r=(m_d-m_u)/(m_s-0.5\cdot(m_d+m_u))$ and
$p_\pi$ and $p_\eta$ are the $\pi$
and $\eta$ momenta in the $\psip$ rest frame. With the   
conventionally accepted values of $m_s=150$ MeV/c$^2$, $m_d=7.5$ MeV/c$^2$,
$m_u=4.2$ MeV/c$^2$
given by Weinberg~\cite{wein}, the ratio $R$ equals 0.0162, which
is smaller than our measurement ($0.048\pm 0.005$)~\cite{note} by a factor of three.
%%%%%%%%%%%%%%%%%%%%%%%%%%%%%%%%%%%%%%%%%%%%%%%%%%%%%%%%%%  
Based on an effective Lagrangian approach, Casalbuoni $et~al$~\cite{aldo}
obtain an improved expression
\beq\label{eq2}
R=\frac{27}{16}\left(\frac{p_\pi}{p_\eta}\right)^3 r^2
\left[\frac{1+\frac{2B}{3A}\frac{\hat{\lambda}f_\pi}
{m^2_{\eta^{\prime}}-m^2_{\pi^0}}}{1+\frac{B}{A}\frac{\hat{\lambda}f_\pi}
{m^2_{\eta^{\prime}}-m^2_\eta}}\right]^2,
\eeq
in which $\hat{\lambda}$ characterizes the
$\eta-\eta^{\prime}$
mixing, $B/A$ is a not yet determined parameter in the effective Lagrangian. $f_\pi=(130\pm 5)$ MeV is obtained from PDG. Using the 
approximation~\cite{yellow}
\beq\label{eq3}  
\hat{\lambda}=\sqrt{\frac{3}{2}}\left(\frac{m^2_{\eta^{\prime}}-m_\eta^2}
{m_s-\frac{m_u+m_d}{2}}\right)\tan\theta_P,
\eeq
where $\theta_P\approx -20^o$ ~\cite{PDG} is the $\eta-\eta^{\prime}$
mixing angle, we obtain $\hat{\lambda}\approx 1.91$ GeV.
With our measured value of $R$, we infer the parameter
$\frac{B}{A}$ equals $-1.42\pm 0.12$ or $-3.11\pm 0.15$ in Equation (~\ref{eq2}).

In terms of QCD multipole expansion, Kuang $et~al$~\cite{kuang}
predict the ratio
\beq\label{eq4}
R^{\prime}
\approx\left(\frac{m_c}{m_b}\right)^2\cdot\left(\frac{p_\eta(\Upsilon^\prime)}
{p_\eta(\psip)}\right)^3\cdot\left(\frac{f(\Upsilon^{\prime})}
{f(\psi(2S))}\right)^2,
\eeq

\beq\label{eq5}  
R^{\prime\prime}\approx\left(\frac{m_c}{m_b}\right)^2\cdot
\left(\frac{p_\eta(\Upsilon^{\prime\prime})}
{p_\eta(\psip)}\right)^3\cdot\left(\frac{f(\Upsilon^{\prime\prime})}
{f(\psi(2S))}\right)^2,
\eeq
where $f(\psi(2S)), f(\Upsilon^{\prime})$, and
$f(\Upsilon^{\prime\prime})$
are the transition amplitudes of $\psi(2S)\ar\jpsi\pi\pi$, $\Upsilon^{\prime}\ar
\Upsilon \pi\pi$, and $\Upsilon^{\prime\prime}\ar
\Upsilon\pi\pi$, respectively, which depend on the potential
model describing the heavy quarkonia. Taking the QCD motivated
Buchm$\ddot{u}$ller-Grunberg-Tye potential
~\cite{bgt} as an example, the predicted values are
$R^{\prime}_{BGT}=0.0025$ and
$R^{\prime\prime}_{BGT}=0.0013$. With our measurements
of ${\cal B}(\psip\ra\Jeta)$
and PDG values of  $\Gamma(\psip)$,
$\Gamma(\Upsilon^{\prime}\ar\Upsilon\eta)$
and
$\Gamma(\Upsilon^{\prime\prime}\ar\Upsilon\eta)$, we obtain
$R^{\prime}_{exp}<0.0098$ and $R^{\prime\prime}_{exp}<0.0065$, which
are consistent with the predictions of Equations (~\ref{eq4}) and
(~\ref{eq5}).

   The BES collaboration acknowledges the staff of BEPC for their hard
efforts. The authors also thank Prof. Y. P. Kuang for enlightening
discussions.
This work is supported in part by the National Natural Science Foundation
of China under contracts Nos. 19991480, 10225524, 10225525, the Chinese
Academy of Sciences under contract No. KJ 95T-03, the 100 Talents
Program of CAS under Contract Nos. U-11, U-24, U-25, and the Knowledge
Innovation Project of CAS under Contract Nos. U-602, U-34 (IHEP);
by the National Natural Science Foundation of China under Contract
No. 10175060 (USTC); and by the Department of Energy under Contract
No. DE-FG03-94ER40833 (U Hawaii).


\begin{thebibliography}{**}

\bibitem{MK1} W. M. Tanenbaum et al., Phys. Rev. Lett. {\bf 36}, 402
                       (1976).

\bibitem{CNTR1} C. J. Biddick et al., Phys. Rev. Lett. {\bf 38}, 1324 (1977).

\bibitem{CNTR2} W. Bartel et al., Phys. Lett. {\bf B79}, 492 (1978).

\bibitem{DASP} R. Brandelik et al., Nucl. Phys. {\bf B160}, 426 (1979).

\bibitem{MK2} T. Himel et al., Phys. Rev. Lett. {\bf 44}, 920 (1980).

\bibitem{CB} M. J. Oreglia et al., Phys. Rev. Lett. {\bf 45}, 959 (1980);
             J. Gaiser et al., Phys. Rev. {\bf D34}, 711 (1986).

\bibitem{Ioffe} B. L. Ioffe and M. A. Shifman, Phys. Lett.  
                                               {\bf B95}, 99 (1980). 

\bibitem{Miller} G. A. Miller et al., Phys. Rep. {\bf 194}, 1 (1990).
\bibitem{aldo}R.casalbuoni $et~al$ Phys. Lett. B309,163(1993).
%\bibitem{ME} K. Gottfried, Phys. Rev. Lett. {\bf 40}, 538 (1978);
%                 M. Voloshin,  Nucl. Phys.  {\bf B154}, 365 (1979);
%                 M. Peskin,  Nucl. Phys.  {\bf B156}, 365 (1979).
%
\bibitem{Yan} T. M. Yan,  Phys. Rev. {\bf D22}, 1652 (1980).
\bibitem{kuang} Y.P.Kuang and T.M.Yan Phys. Rev. D24,2874(1981);\\
                Y.P.Kuang $et~al$ Phys. Rev. D37,1210(1988).
%\bibitem{Voloshin} M. Voloshin and V. Zakharov,  Phys. Rev. Lett. {\bf 
%           45}, 688 (1981).
%          ; M. Shifman, Phys. Rep. {\bf 209}, 341 (1991).

\bibitem{bes} J. Z. Bai et al., BES Collab., Nucl. Instr. Meth.
              {\bf A344}, 319 (1994).

\bibitem{bes2} J. Z. Bai et al., BES Collab., Nucl. Instr. Meth.
              {\bf A458}, 627 (2001).

\bibitem{Npsip} X. H. Mo et al., accepted to HEP \& NP.

\bibitem{PDG} Particle Data Group, K. Hagiwara et al., Phys. Rev.
 {\bf D66}, 010001 (2002).
   
\bibitem{R} J. Z. Bai et al., BES Collab., Phys. Rev. Lett.
                              {\bf 84}, 594 (2000).

\bibitem{gener} J. Z. Bai et al., BES Collab., Phys. Rev. 
                              {\bf D62}, 032002 (2000).

\bibitem{photons}  J. Z. Bai et al., BES Collab., submitted to Phys. Rev. D, hep-ex/0402013.

\bibitem{lambda} J. Z. Bai et al., BES Collab., Phys. Rev. {\bf D67},
  112001 (2003).

\bibitem{2003pdg} The branching fractions for $\chi_{c1,2}\ar\gamma\jpsi$
are quoted from the 2003 WWW update of the Review of Particle Physics
(http://pdg.lbl.gov/2003/mxxx.html).

\bibitem{wein} S. Weinberg,  ``The problem of mass'', in a Festschrift for 
   I. I. Rabi, ed. L. Motz (New York Academy of Sciences, New York, 1978).

%\bibitem{Ntot} J. Z. Bai et al., BES Collab., Phys. Lett.
%                              {\bf B550}, 24 (2002).

\bibitem{note} Part of the systematic error cancels in the calculation
  of $R$.

\bibitem{yellow}Y.P.Kuang, private Communications; see also CERN Yellow 
   Report on $Quarkonium~Physics$ (to be published).

\bibitem{bgt} W.Buchm$\ddot{u}$ller $et~al$. Phys. Rev. Lett.45,103(1980);\\
              W.Buchm$\ddot{u}$ller $et~al$. Phys. Rev. D24,132(1981).

\end{thebibliography}
\end{document}